# Mathematical Modeling of Heat Conduction

**Abdul Aziz Momin**[1] , **Nikhil Shende**[1], **Abhijna Anamtatmakula**[1], **Emily Ganguly**[1], **Ashwin Gurbani**[1], **Chaitanya A Joshi**[2], **Yogesh Y Mahajan**[3]

1. Department of Metallurgical and Materials Engineering (MME), Visvesvaraya National Institute of Technology(VNIT), Nagpur, India
2. Assistant Professor, Metallurgy and Materials Department, VNIT Nagpur
3. Associate Professor, Metallurgy and Materials Department, VNIT Nagpur

**Abstract**- This report describes a mathematical model of heat conduction. The differential equation for heat conduction in one dimensional rod has been derived. The explicit finite difference numerical method is used to solve this differential equation. Then for simulation, a code was written in using python libraries via Jupyter notebook. The simulation carried out for Aluminum, Copper and Mild Steel rods and results were discussed.

## 1. Introduction:

An Equation that contains both variables and derivatives of one of the or more than one variable, it is called as a Differential Equation (DE). To solve such an equation, it must be divided into simpler parts and then solved accordingly. [1] The figure.1 shows the how to apply or use DE in real world problems.

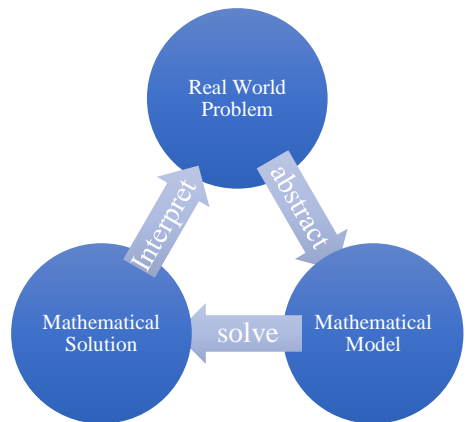

Figure.1 : Differential Equations

In applications, the functions generally represent physical quantities, the derivatives represent their rates of change, and the differential equation defines a relationship between the two. Such relations are common; therefore, differential equations play a prominent role in many disciplines including engineering, physics, economics, and biology, etc.

In day to day life DEs are used in many scientific and analytical research. Classifications of differential equations are done as per following chart (See Table 1). Based on types differential equations are classified into 2 categories: [2,3]

- Ordinary Differential Equation
- Partial Differential Equation

Table.1 : Classification of Differential Equations

| Sr.No. | Classification | Types |
|---|---|---|
| 1. | Based on Type | 1. Ordinary<br>2. Partial |

| 2. | Based on Order | 1. 1st Order |
| | | 2. 2nd and Nth Order |
| 3. | Based on Linearity | 1. Linear |
| | | 2. Non Linear |
| 4. | Based on Homogeneity | 1. Homogeneous |
| | | 2. Non Homogeneous |

**2. Heat Equation:**
In mathematics and physics, the heat equation is a certain PDE. Solutions of the heat equation are sometimes known as caloric functions.
The heat equation describes the distribution of heat in at time t. The heat equation –
$$\partial f/\partial t = \partial^2 f/\partial x^2$$

2.1 Partial Differential Equation[4]:
The heat equation is a partial differential equation (PDE) –
$$\partial T/\partial t = \alpha(\partial^2 T/\partial x^2)$$

Where $\alpha$ is the diffusion coefficient. Assume the initial distribution is a spike at $x = 0$ and is zero elsewhere. The finite difference method is shown in figure 2.

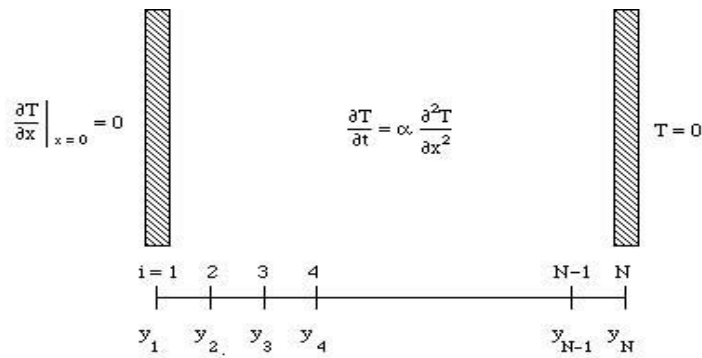

Figure.2 : Finite difference method.

2.2 Solution of 1-D Heat Equation:
While there are many methods to solve the 1D heat equation like Implicit method, Runge Kutta method, explicit method, the explicit method is chosen in this project as the explicit method is especially well-suited to solving high-speed dynamic events that require many small increments to obtain a high-resolution solution. If the duration of the event is short, the solution can be obtained efficiently.

For solving the 1D heat equation in explicit method, we will require the 1D heat equation, the values of temperature at all spatial points at time=0, and the values of temperature at the ends of the rods. Namely,
$$\alpha \frac{\partial^2 T}{\partial x^2} = \frac{\partial T}{\partial t} \text{[5]}$$
$$T(0, t) = T_1$$
$$T(L, t) = T_2$$
$$T(L/2, t) = T_b$$

Where, $T_1, T_2$ and $T_b$ are the temperatures at initial, final and mid point. To proceed further, the length of the rod is divided into spaces of length $\Delta x = \frac{L}{n}$, creating n+1 nodes including n=0. The time is similarly broken into time steps of $\Delta t$. Hence, temperature $T_i^j$ corresponds to the temperature at node i, that is, $x=(i)(\Delta x)$ and $t=(j)(\Delta t)$. If, time step= j, space step= i, i-1, i+1, then the temperature at time step= j+1 and space step=i can be found.
Using Central divided difference approximation to the left hand side (x dependent part) of the heat equation at a general interior point (i,j),

$$\frac{\partial^2 T}{\partial x^2} = \frac{T_{i+1}^j - 2T_i^j + T_{i-1}^j}{(\Delta x)^2}$$

Again using forward divided difference approximation to the right hand side (time dependent part) of the heat equation at the same interior point (i,j),

$$\left.\frac{\partial T}{\partial t}\right|_{i,j} = \frac{T_i^{j+1} - T_i^j}{\Delta t}$$

Substituting these approximations into the governing equation,

$$\alpha \frac{\partial^2 T}{\partial x^2} = \frac{\partial T}{\partial t}$$

$$\alpha \frac{T_{i+1}^j - 2T_i^j + T_{i-1}^j}{(\Delta x)^2} = \frac{T_i^{j+1} - T_i^j}{\Delta t}$$

$$T_i^{j+1} = T_i^j + \alpha \frac{\Delta t}{(\Delta x)^2}(T_{i+1}^j - 2T_i^j + T_{i-1}^j)$$

$$\lambda = \alpha \frac{\Delta t}{(\Delta x)^2}$$

$$T_i^{j+1} = T_i^j + \lambda(T_{i+1}^j - 2T_i^j + T_{i-1}^j)$$

Hence, if the temperatures at time step= j, space step= i, i-1, i+1,then the temperature at time step= j+1 and space step=i can be found. This is the final equation, which is used for the simulation.

**3.Implementation of Code and Simulation:** [6]

Before writing a code for the simulation of the mathematical model, boundary conditions need to be defined –

3.1 Boundary condition: [7]

In mathematics, in the field of differential equations, a boundary value problem is a differential equation together with a set of additional constraints, called the boundary conditions. A solution to a boundary value problem is a solution to the differential equation which also satisfies the boundary conditions.

There are various types of boundary conditions. In this project two types of boundary conditions used. Dirichlet and Neumann boundary conditions.

There are several types of boundary conditions, in table 2 two conditions have been discussed, Neumann and Dirichlet boundary conditions.

Table.2 : Types of Boundary Conditions

| Sr.No. | Types | Condition |
|---|---|---|
| 1. | Neumann Condition | 1. y" + y = 0<br>2. y'(0)=0 & y'(1)=0 |
| 2. | Dirichlet Condition | 1. y" + y =0<br>2. y(0)=0 & y(1)=1 |

3.1.1 Dirichlet Condition:

The mathematical expression for Dirichlet condition is given in the table 3. In other words[7], the Dirichlet boundary condition state the value that the solution function f to the differential equation must have on the boundary of the domain C. The boundary is usually denoted as ∂C. In a two-dimensional domain that is described by *x* and *y*, a typical Dirichlet boundary condition would be -

    f (x,y) = g (x,y,...), where : (x,y) ∈ ∂C

Here the function *g* may not only depend on *x* and *y*, but also on additional independent variables, e.g. , the time t. Our example of the barrier lake used a Dirichlet boundary condition stating that the volume of the lake was 0 for t = 0. Here the function *g* is a constant, but it must not necessarily be the case.

### 3.1.2 Neumann Condition:
The mathematical expression for Neumann condition is given in the table 3. In other words[7], the second important type of boundary conditions are Neumann boundary conditions. Neumann boundary conditions state that the derivative of the solution function f to the differential equation must have a given value on the boundary of the domain C. A typical Neumann boundary condition would be -

$$\partial f(x,y) / \partial x = g(x,y,...) \quad ; \text{where} : (x,y) \in \partial C \quad \text{OR}$$
$$\partial f(x,y) / \partial y = g(x,y,...) \quad ; \text{where} : (x,y) \in \partial C$$

Again the function *g* may not only depend on *x* and *y*, but also on additional variables such as time.

### 3.2 Code:
In this section, 1D simulation[8,9] is presented of the subsequent mathematical model. The code is fairly short, since the explicit methods are nice and elegant. In the beginning download the python libraries such as numpy, Matplotlib, axes 3-d, scipy, and mpl_toolkit. [10,11,12]

```
#
1    import numpy as np
2    from scipy.integrate import odeint
3    import matplotlib.pyplot as plt
4    N = 100    # number of points to discretize
5    L = 100.0
6    X = np.linspace(0, L, N) # position along the rod
7    h = L / (N - 1)
8    α = 3.98        #  λ = ((K * t) / (rho * C))
9    def odefunc(u, t):
10       dudt = np.zeros(X.shape)
11          dudt[0] = 0 # constant at boundary condition
12          dudt[-1] = 0
13   # now for the internal nodes
14   # Let rho is density of material and C is specific heat capacity
15   for i in range(1, N-1):
16       dudt[i] = ((α * t) / (rho * C)) *  (u[i + 1] - 2*u[i] + u[i - 1]) / h**2
17   return dudt
18   init = 0.0 * np.ones(X.shape) # initial temperature
19   init[0] = 0.0  # one boundary condition
20   init[50] = 50.0
21   init[-1] = 0.0 # the other boundary condition
22   tspan = np.linspace(0.0, 6000.0, 100)
23   sol = odeint(odefunc, init, tspan)
24   for i in range(0, len(tspan), 5):
25       plt.plot(X, sol[i], label='t={0:1.2f}'.format(tspan[i]))
26   # put legend outside the figure
27       plt.legend(loc='center left', bbox_to_anchor=(1, 0.5))
28       plt.xlabel('X position')
29       plt.ylabel('Temperature')
30   # adjust figure edges so the legend is in the figure
31       plt.subplots_adjust(top=0.89, right=0.77)
32   #plt.savefig('images/pde-transient-heat-1.png')
33   # Make a 3d figure
34   from mpl_toolkits.mplot3d import Axes3D
35   fig = plt.figure()
36   ax = fig.add_subplot(111, projection='3d')
```

```
37    SX, ST = np.meshgrid(X, tspan)
38    ax.plot_surface(SX, ST, sol, cmap='jet')
39    ax.set_xlabel('X')
40    ax.set_ylabel('time')
41    ax.set_zlabel('T')
42    ax.view_init(elev=15, azim=-124)
43    plt.show()
44    # adjust view so it is easy to see
45    #plt.savefig('images/pde-transient-heat-3d.png')
46    # animated solution. We will use imagemagick for this
47    # we save each frame as an image, and use the imagemagick convert command to
48    # make an animated gif
49    for i in range(len(tspan)):
50        plt.clf()
51        plt.plot(X, sol[i])
52        plt.xlabel('X')
53        plt.ylabel('T(X)')
54        plt.title('t = {0}'.format(tspan[i]))
55    plt.show()
```

**4. Results and Inference:**

The results will be different for different boundary conditions. Subsequent results are shown below for different rods for different conditions. In the table 3, the thermal properties of these are rods(at 50°C) are given –

Table.3: Thermal Properties of different rods.

| Rod/Propeties | Thermal Conductivity($\alpha$)(W/m-K)[13] | Density(g/cc)[14] | Specific heat capacity(g-°C)[15] |
|---|---|---|---|
| Aluminium (Al) | 2.38 | 2.7 | 0.92 |
| Copper (Cu) | 4.1 | 8.96 | 0.376 |
| Mild Steel | 0.064 | 7.85 | 0.51 |

4.1 For Neumann Condition:

The both end of the rod is kept at 0°C. While the temperature of the other locations kept changing. For Neumann condition, apply the Neumann boundary condition in the code. In the line number 11 and 12, put boundary conditions as,
dudt[0] = 0  &  dudt[1] = 0.
For different rod, we need to change the thermal properties, hence;
1) In the line number 8, change the thermal conductivity value.
2) In the line number 16, put density and specific heat capacity value of different materials.

4.1.1 For Aluminium Rod:

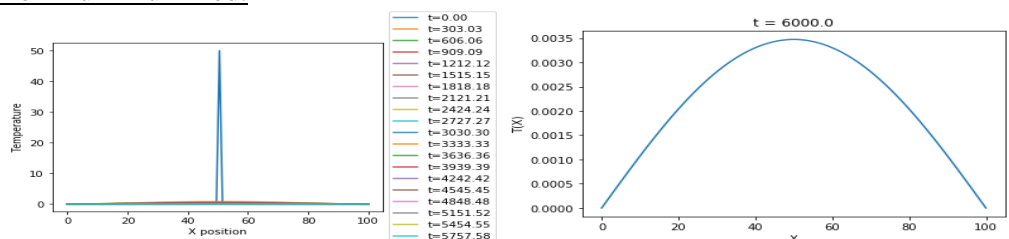

Figure.3 : a) Temperature profile at different time ;  b) Temp. vs X vs time at steady state.

The sharp peak in figure 3(a) shows the temperature profile at t=0 sec. where mid point is at 50°C and all other points is at 0°C. The figure 3(b) shows the steady state temperature profile. The both ends are at 0°C where as temperature at all other points kept changing, its due to Neumann boundary condition.

4.1.2 For Copper Rod:

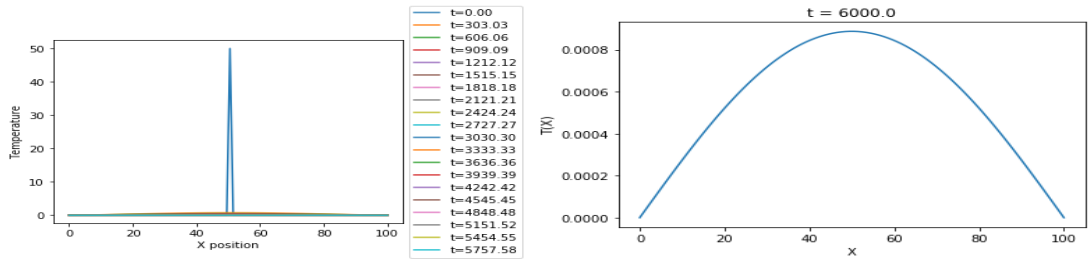

Figure.4 : a) Temperature profile at different time ; b) Temp. vs X vs time at steady state.

The sharp peak in figure 4(a) shows the temperature profile at t=0 sec. where mid point is at 50°C and all other points is at 0°C. The figure 4(b) shows the steady state temperature profile. The both ends are at 0°C where as temperature at all other points kept changing, its due to Neumann boundary condition. However, the rate of heat conduction is more as compare to Al rod due to the higher value of thermal conductivity of copper.

### 4.1.3 For Mild Steel Rod:

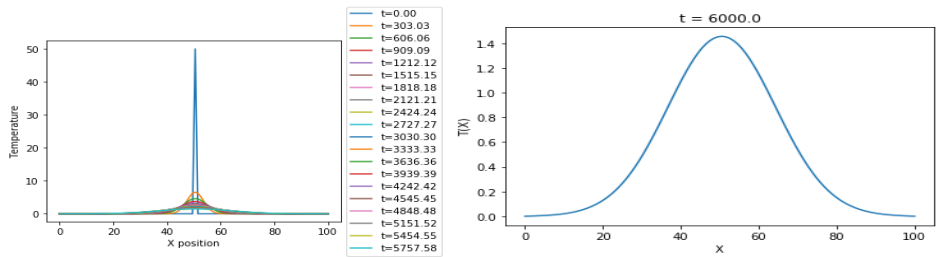

Figure.5 : a) Temperature profile at different time ; b) Temp. vs X vs time at steady state.

The sharp peak in figure 5(a) shows the temperature profile at t=0 sec. where mid point is at 50°C and all other points is at 0°C. The figure 3(b) shows the steady state temperature profile. However, the rate of heat conduction is less as compare to both Al and Cu rod due to the lesser value of thermal conductivity of mild steel. After very long time(6000 sec.) it has reached to a steady state with both the ends are at 0°C because of Neumann boundary condition.

### 4.2 For Dirichlet Condition:

The one end of the rod is kept at 0°C, and at all other locations the temperature kept changing with time. For Dirichlet condition, apply the Dirichlet boundary condition in the code. In the line number 11 and 12, put boundary conditions as, u[0] = 0  &  u[1] = 50.

For different rod, we need to change the thermal properties, hence;

1) In the line number 8, change the thermal conductivity value.
2) In the line number 16, put density and specific heat capacity value of different materials.

### 4.2.1 For Aluminium Rod:

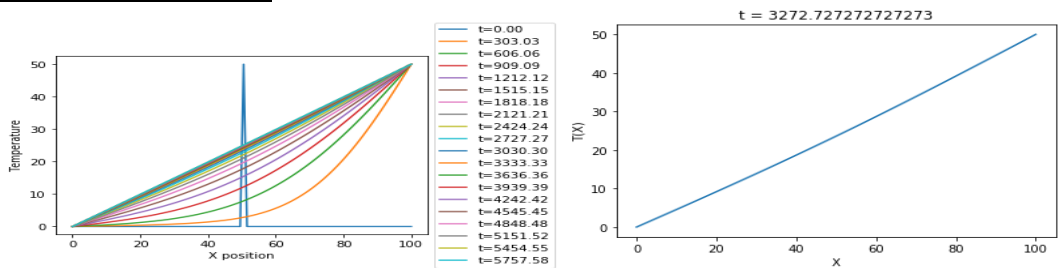

Figure.6 : a) Temperature profile at different time ; b) Temp. vs X vs time at steady state.

The sharp peak in figure 6(a) shows the temperature profile at t=0 sec. where mid point is at 50°C and all other points is at 0°C. Figure 6(b) shows the steady state temperature profile. After very long time(3272 sec.) it has reached to a steady state with one end is at 0°C and the other end has reached to maximum 50°C because Dirichlet boundary condition is used.

### 4.2.2 For Copper Rod:

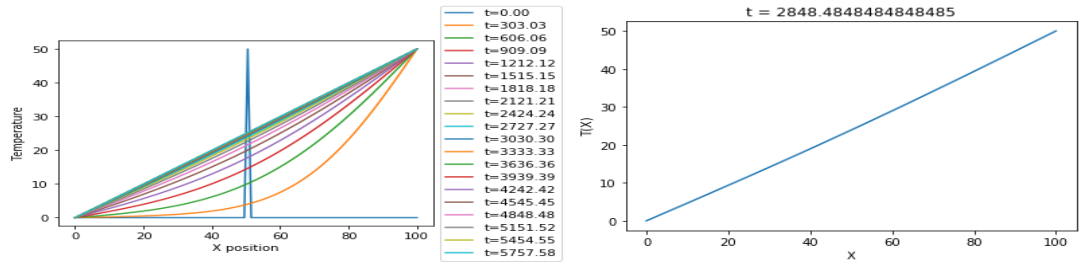

Figure.7 : a) Temperature profile at different time ; b) Temp. vs X vs time at steady state.

The sharp peak in figure 7(a) shows the temperature profile at t=0 sec. where mid point is at 50°C and all other points is at 0°C. Figure 7(b) shows the steady state temperature profile. However, the rate of heat conduction is more as compare to Al rod due to the higher value of thermal conductivity of copper. After very long time(2848 sec.) it has reached to a steady state with one end is at 0°C while the other end has rised upto maximum 50°C as because Dirichlet boundary condition is used.

4.2.3 For Mild Steel Rod:

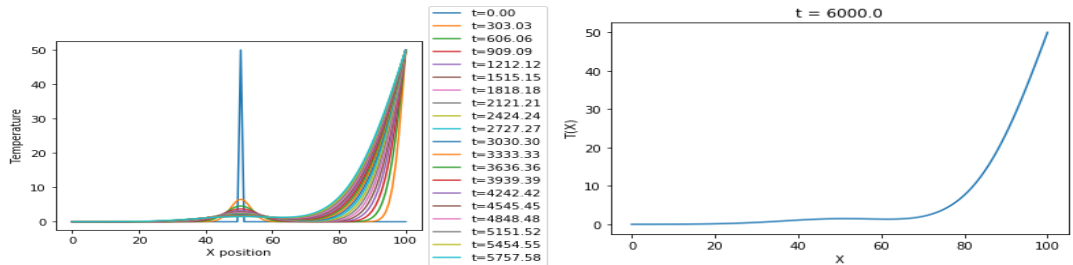

Figure.8 : a) Temperature profile at different time ; b) Temp. vs X vs time at steady state.

The sharp peak in Figure.18 a shows the temperature profile at t=0 sec. where mid point is at 50°C and all other points is at 0°C. Figure 6(b) shows the steady state temperature profile. However, the rate of heat conduction is less as compare to both Al and Cu rod due to the lesser value of thermal conductivity of mild steel. After very long time(6000 sec.) it has reached to a steady state with one end is at 0°C while the other end has rised upto maximum 50°C because of Dirichlet boundary condition.

4.2.4 Comparison of time to reach the steady state:

Due to the difference in the thermal conductivity values(given in table 3) of Al, Cu and Mild Steel rods, the time required to reach the steady state is different. The time required for mild steel rod is very large as compared to Al and Cu rods, whereas this difference is negligible for later rods due to less difference in thermal conductivity.

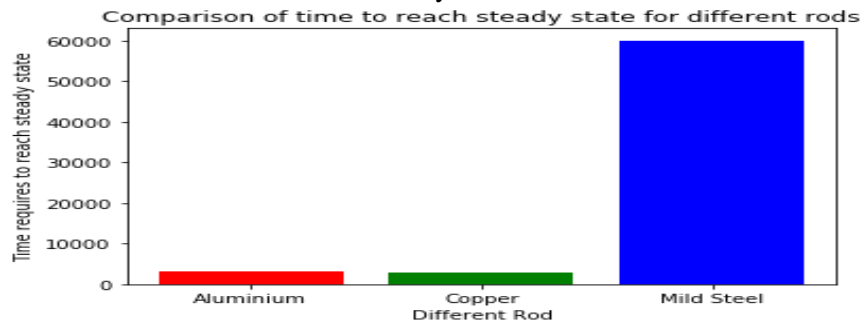

Figure.9 : Comparison of time to reach the steady state for different rods.

4.3 Inference:

The difference between Neumann and Dirichlet boundary condition is the constant temperatures at the end nodes. Hence, different curvatures, the parabolic curve in Neumann condition while linear curve in Dirichlet condition at the steady state. This is due to constant value of DE, in Neumann the first derivate at node is constant while in Dirichlet the DE at node itself is constant.

The temperature profiles are given above. The difference in these temperature profiles is due to -
- The difference in the slope of curve is due to difference in material's property (thermal conductivity α, density rho and specific heat capacity C).

- Due to difference in material's property the time require to reach the steady state is different.
- This difference is due to different value of thermal diffusivity.

**5. Conclusion:**

This is the mathematical model for heat conduction in 1-dimension. The subsequent results and inferences have drawn above. In this mode –

1. The simulations are carried out for different rods for Neumann and Dirichlet conditions.
2. This mathematical model can be used for any material for 1-D heat conduction.
3. In the future, more work can be done for 2-D and 3-D applications.

**6. References:**

1. Research on a Class of Ordinary Differential Equations and Application in Metallurgy : DOI: 10.1007/978-3-642-16339-5_52
2. Types of solutions of differential equations : DOI: 10.1007/978-3-7643-8638-2_7
3. Differential Equations, Partial : DOI: 10.1016/B0-12-227410-5/00173-3
4. Articles in Heat transfer research- Jan 2017; DOI:10.1615/HeatTransRes.2017017295
5. Differential Equations, Partial : DOI: 10.1016/B0-12-227410-5/00173-3
6. A study on an analytic solution 1D heat equation of a parabolic partial differential equation and implement in computer programming : International Journal of Scientific & Engineering Research Volume 9, Issue 9, September-2018; ISSN 2229-5518
7. Engineering Mathematics : DOI:10.1016/B978-1-4557-3141-1.50003-4
8. 600.112: Intro Programming for Scientists and Engineers
9. Steps: modeling and simulating complex reaction-diffusion systems with Python; DOI: 10.3389/neuro.11.015.2009
10. Heat simulation via Scilab programming : DOI: 10.1063/1.488757
11. Numerical Simulation of one dimensional Heat Equation: B-Spline Finite Element Method: Internation Journal of Computer Science and Engineering: (IJSCSE), Vol.2 No.2, Apr-May 2011, ISSN: 0976-5166
12. Steps: modeling and simulating complex reaction-diffusion systems with Python: DOI: 10.3389/neuro.11.015.2009
13. Mokhena, T. C., Mochane, M. J., Sefadi, J. S., Motloung, S. V., & Andala, D. M. (2018). Thermal Conductivity of Graphite-Based Polymer Composites. Impact of Thermal Conductivity on Energy Technologies. doi:10.5772/intechopen.75676
14. Beer.F.P. , Johnston.E.R. (1992). Mechanics of Materials , 2nd edition. McGraw-Hill
15. J. Carvill, in Mechanical Engineer's Data Handbook, 1993